\pdfoutput=1
\documentclass[galaxies,article,accept,moreauthors,pdftex,10pt,a4paper]{mdpi}
\usepackage{float}
\usepackage[labelformat=simple]{subfig}

\usepackage{upgreek}
\usepackage{textcomp}

\firstpage{1}
\articlenumber{50}
\doinum{10.3390/galaxies6020050}
\pubvolume{6}
\pubyear{2018}
\copyrightyear{2018}
\history{Received: 16 February 2018; Accepted: 19 April 2018; Published: 24 April 2018}

\Title{A Study of Background Conditions for Sphinx---The Satellite-Borne Gamma-Ray Burst Polarimeter}

\Author{Fei Xie $^{1,2,}$*\orcidA{} and Mark Pearce $^{1,2,\dagger}$\orcidB{}} 

\address{%
$^{1}$ \quad Department of Physics, KTH Royal Institute of Technology, 106 91 Stockholm, Sweden\\
$^{2}$ \quad The Oskar Klein Centre for Cosmoparticle Physics, AlbaNova University Centre, 106 91 Stockholm, Sweden}
\corres{Correspondence: fxie@kth.se}
\firstnote{On behalf of the SPHiNX Collaboration.}

\abstract{\textcolor{black}{SPHiNX is a proposed satellite-borne gamma-ray burst polarimeter operating in the energy range 50--500 keV.} The mission aims to probe the fundamental mechanism responsible for gamma-ray burst prompt emission through polarisation measurements. Optimising the signal-to-background ratio for SPHiNX is an important task during the design phase. The Geant4 Monte Carlo toolkit is used in this work. From the simulation, the total background outside the South Atlantic Anomaly (SAA)
is about 323 counts/s, which is dominated by the cosmic X-ray background and albedo gamma rays, which contribute $\sim$60\% and $\sim$35\% of the total background, respectively. The background from albedo neutrons and primary and secondary cosmic rays is negligible. The delayed background induced by the SAA-trapped protons is about 190 counts/s when SPHiNX operates in orbit for one year. The~resulting total background level of $\sim$513 counts/s allows the polarisation of \mbox{$\sim$50 GRBs}
with minimum detectable polarisation less than 30\% to be determined during the two-year mission~lifetime.}

\keyword{polarimeter; Compton scattering; GRB; background}

\begin{document}

\section{Introduction}
\label{sec:intro}
The Satellite Polarimeter for High eNergy X-rays (SPHiNX) is a proposed mission for a Swedish scientific satellite based on the InnoSat platform\footnote{InnoSat System Requirements Document from OHB-Sweden, \url{http://www.snsb.se/Global/Forskare/Utlysningar/InnoSat\%20System\%20Requirements\%20Document_IS-OSE-RS-0001_2C.pdf}}, which supports a maximum payload mass of 25 kg and provides a payload power budget of 30 W.

SPHiNX is a dedicated polarimeter for gamma-ray bursts (GRBs), the most luminous explosions in the Universe~\cite{ref:grb}. Long GRBs are generated by the collapse of massive stars~\cite{ref:collapse}, while short GRBs come from the merger of binary neutron stars or neutron star--black hole binary systems~\cite{ref:merge}. The~recent detection of gravitational waves together with a short-duration GRB, from a binary neutron star system~\cite{ref:gw}, highlights the importance of GRB measurements. Even though thousands of GRBs have been detected by, for example, the CGRO, Swift and Fermi\footnote{FERMIGBRST---Fermi GBM Burst Catalog, which has been used in the performance simulation of SPHiNX, \url{https://heasarc.gsfc.nasa.gov/W3Browse/fermi/fermigbrst.html}} satellites, there are still many open questions about~GRBs.

The main scientific goal of the SPHiNX mission is to probe the fundamental mechanism behind the GRB prompt emission, by measuring the linear polarisation~\cite{ref:grbmodel}. In contrast with energy spectrum and timing studies, there is a lack of reliable polarimetric data.

Background plays an important role in the satellite-borne high-energy telescopes as they operate in the severe radiation environment above the atmosphere. Particles in the orbital environment interact with the detector, \textcolor{black}{generating background indistinguishable from the signal data,} resulting in a loss of sensitivity.
The effect of background particles can be simulated with Monte Carlo software. Such~simulations provide an efficient method to optimise the scientific performance of an instrument, for example, by maximising the signal-to-background ratio (S/B)~\cite{ref:Dean}.

This paper describes the background study for SPHiNX implemented with the Geant4~\cite{ref:geant4} Monte Carlo simulation toolkit, and the paper is organised as follows. \textcolor{black}{Section~\ref{sec:design} provides an overview of the SPHiNX mission and its properties.} Section~\ref{sec:g4sim} presents the details of the \textcolor{black}{simulations}, and Section~\ref{sec:results} presents results from the \textcolor{black}{simulations}. Section~\ref{sec:summary} is a brief summary of the work.

\section{Instrument Design and Properties}
\label{sec:design}

SPHiNX is optimised for GRB detection, meaning that a large field of view (FoV) is needed. Compton scattering is the dominating physics process in the SPHiNX sensitive energy range (50--500~keV). The photon interaction cross-section is described by the Klein--Nishina equation as~follows:
\begin{equation}
\frac{d\sigma}{d\Omega} = \frac{r_0^2 \varepsilon^2}{2} (\frac{1}{\varepsilon}+\varepsilon- 2sin^2\theta cos^2 \psi),
\label{equ:klein}
\end{equation}
where $r_0$ is the classical electron radius, $\varepsilon = E'/ E$, $E$ is the initial photon energy, $E'$ is the scattered photon energy, $\theta$ is the polar scattering angle between the incident photon and the scattered photon, and $\psi$ is the azimuthal scattering angle of the scattered photon with respect to the polarisation of the incident photon.

From Equation~(\ref{equ:klein}), it is clear that the azimuthal scattering angle $\psi$ will be modulated by the polarisation of the incident photons, that is, photons are preferentially scattered at angles perpendicular to the incident polarisation vector. This mechanism provides a sinusoidal modulation curve when reconstructing the azimuthal scattering angle recorded by the detector. The phase and amplitude of the modulation curve are proportional to the polarisation angle and fraction of the incident beam, respectively.

\begin{figure}[H]
\centering
\includegraphics[width=0.8\textwidth]{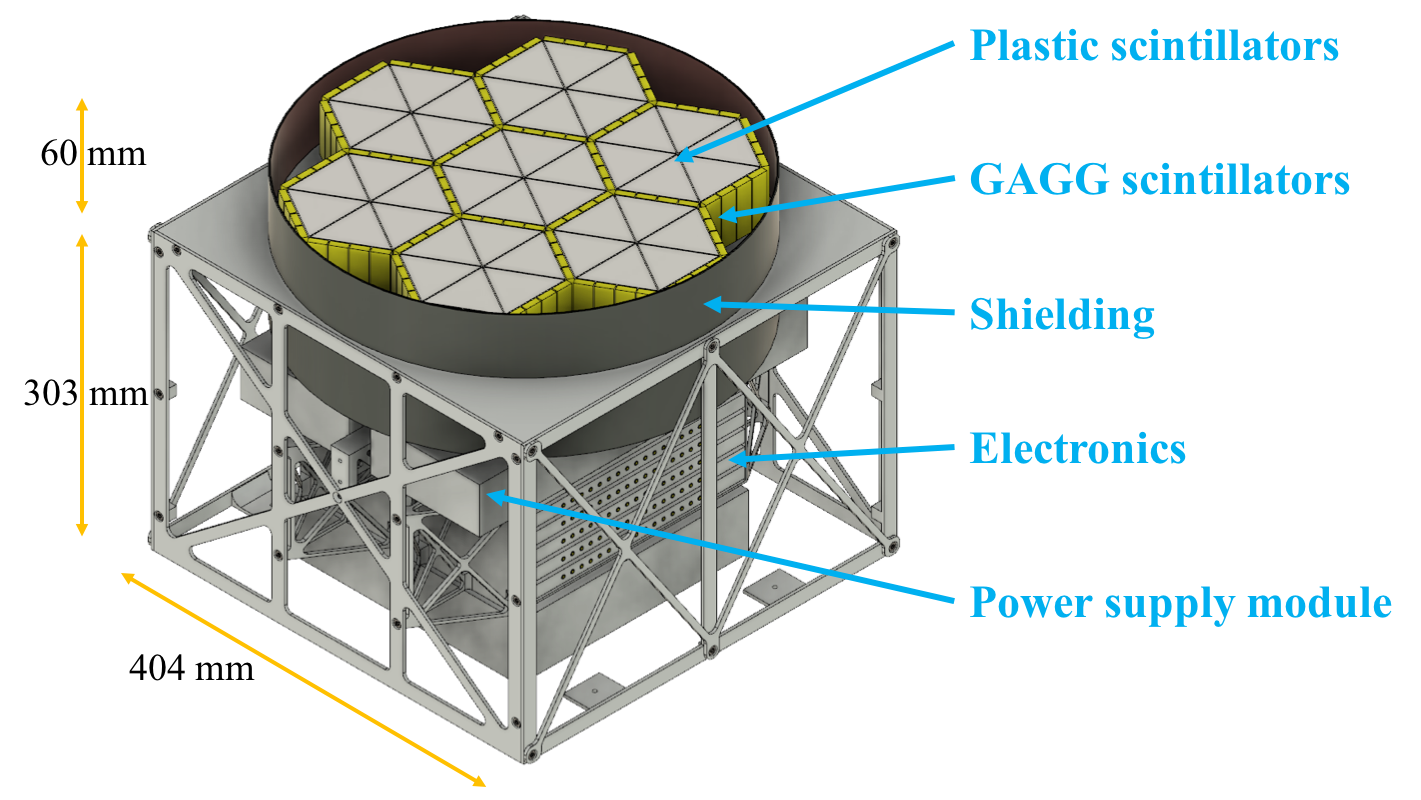}
\caption{An overview of the SPHiNX payload, including the plastic and GAGG scintillators, the~shielding, electronics, and the power-supply module.} 
\label{fig:payload}
\end{figure}

\textcolor{black}{Unlike the GRB polarimeter, POLAR~\cite{ref:polar}, which makes measurements on-board the Chinese spacelab Tiangong-2, two kinds of detector material have been used for SPHiNX;} one is the low-atomic-number plastic scintillator which has a large cross-section for Compton scattering, and the other one is the high-atomic-number GAGG (gadolinium aluminium gallium garnet) scintillator\textcolor{black}{~\cite{ref:gagg}}, which provides a high probability for photoelectric absorption. In the baseline design, plastic scintillators are read out using photomultiplier tubes (PMTs), and GAGG scintillators are read out using multipixel photon counters (MPPCs).  As shown in Figure~\ref{fig:payload}, inside the cylindrical shielding, seven hexagonal plastic scintillators are indicated in grey, and each of them is split into six pieces with a gap size of 1~mm in between. The gap is implemented to accommodate wrapping materials that are needed to maximise the collection of scintillation light. Each side of the plastic scintillators is surrounded by four pieces of GAGG scintillator which are shown in yellow, resulting in a total number of 120 pieces of~GAGG.

\textcolor{black}{The ideal events for polarisation measurement are generated by photons that undergo Compton scattering in a plastic scintillator with a subsequent photoelectric absorption in a GAGG scintillator. The interactions are required to occur within a coincidence window with duration of a few hundred nanoseconds.} In practice, all valid two-hit events can be used for the polarisation calculation, that is, plastic to plastic, GAGG to GAGG, plastic to GAGG and GAGG to plastic.  One-hit events are foreseen to be used for spectroscopy and localisation of GRBs.
Properties of SPHiNX are shown in Table~\ref{table:property}. \textcolor{black}{The~field of view is defined by the incidence angle for which the effective area reduces to 50\% of the value for an on-axis observation.}

\unskip
\begin{table}[H]
\caption{SPHiNX properties.}
\centering
\begin{tabular}{cc}
\toprule
\textbf{Property}	& \textbf{Value}\\
\midrule
Field of view		& 	$\pm$60$^{\circ}$		\\
Energy range		& 	50--500~keV			\\
Effective area		& 	$\geq$125 cm$^2$ 	\\
Energy resolution	& 	$\leq$30\% @ 60~keV	\\
Timing resolution	& 	$\leq$100 ms			\\
\bottomrule
\end{tabular}
\label{table:property}
\end{table}

\section{Geant4 Simulation}
\label{sec:g4sim}

Geant4~\cite{ref:geant4} is a powerful toolkit for Monte Carlo simulations of particle interactions.
The geometric mass model of the satellite, the input particle spectrum, physics models describing particle interactions and the selections on energy deposits are needed in order to determine background rates.
Geant4 \textcolor{black}{version 10.02.p02\footnote{\url{http://geant4.web.cern.ch/support/download_archive?page=1}}} has been used in this work.

\subsection{Geometric Model}
\begin{figure}[H]
\centering
\includegraphics[width=0.45\textwidth]{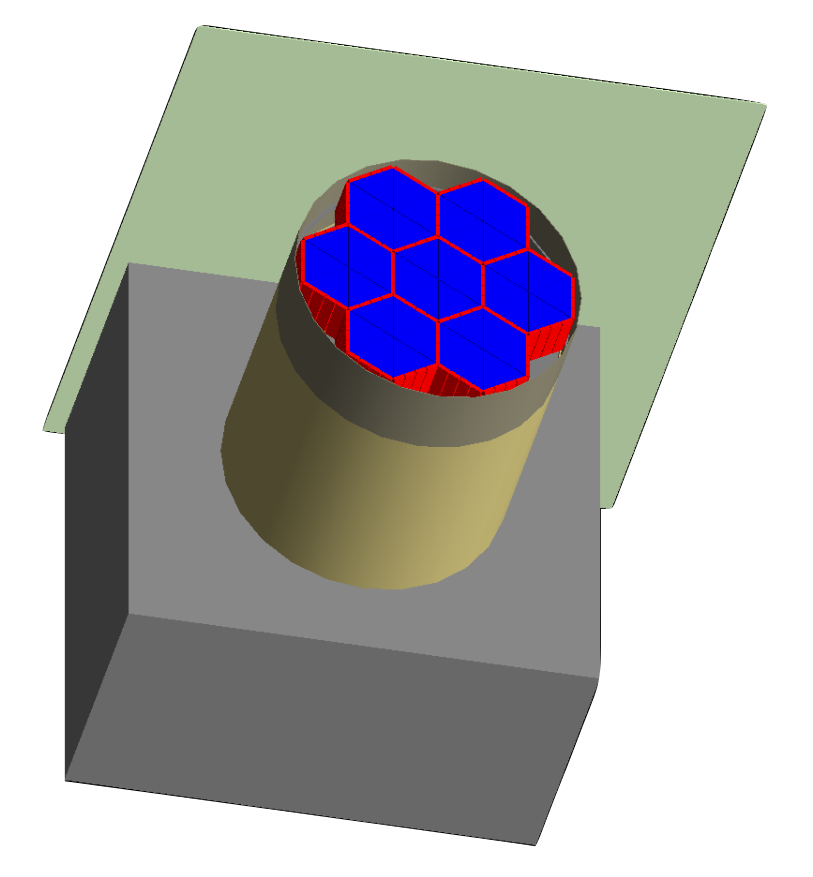}
\caption{The SPHiNX satellite as implemented in Geant4. The figure shows the geometry of the solar panel (green), InnoSat platform (grey), shielding (the upper brown cylinder), support structure (the~bottom yellow cylinder) and sensitive scintillators (blue and red components inside the shielding). For~clarity, the CFRP window covering the instrument aperture is not shown.} 
\label{fig:g4payload}
\end{figure}

A geometric model of the SPHiNX satellite is built in Geant4 based on the mission baseline design. All important components are implemented, including the solar panel, scintillator detectors, photosensors, carbon fibre-reinforced polymer (CFRP) front window, electronics, shielding, support structure, and the InnoSat platform, as shown in Figure~\ref{fig:g4payload}. A higher level of detail is implemented in the sensitive detectors and elements around them, for example, the cylindrical shielding comprising three layers of 1~mm of lead (outermost layer), 0.5~mm of tin (intermediate layer) and 0.25~mm of copper (innermost layer). \textcolor{black}{For components with a complicated structure that lies far from the sensitive detector, the primary concern is to represent the correct mass and the dominant materials of the modelled objects. For example, for the InnoSat platform, structural 7075 aluminium alloy and the lithium battery have been implemented.}

\subsection{Space Radiation Environment}
\begin{figure}[H]
\centering
\includegraphics[width=0.7\textwidth]{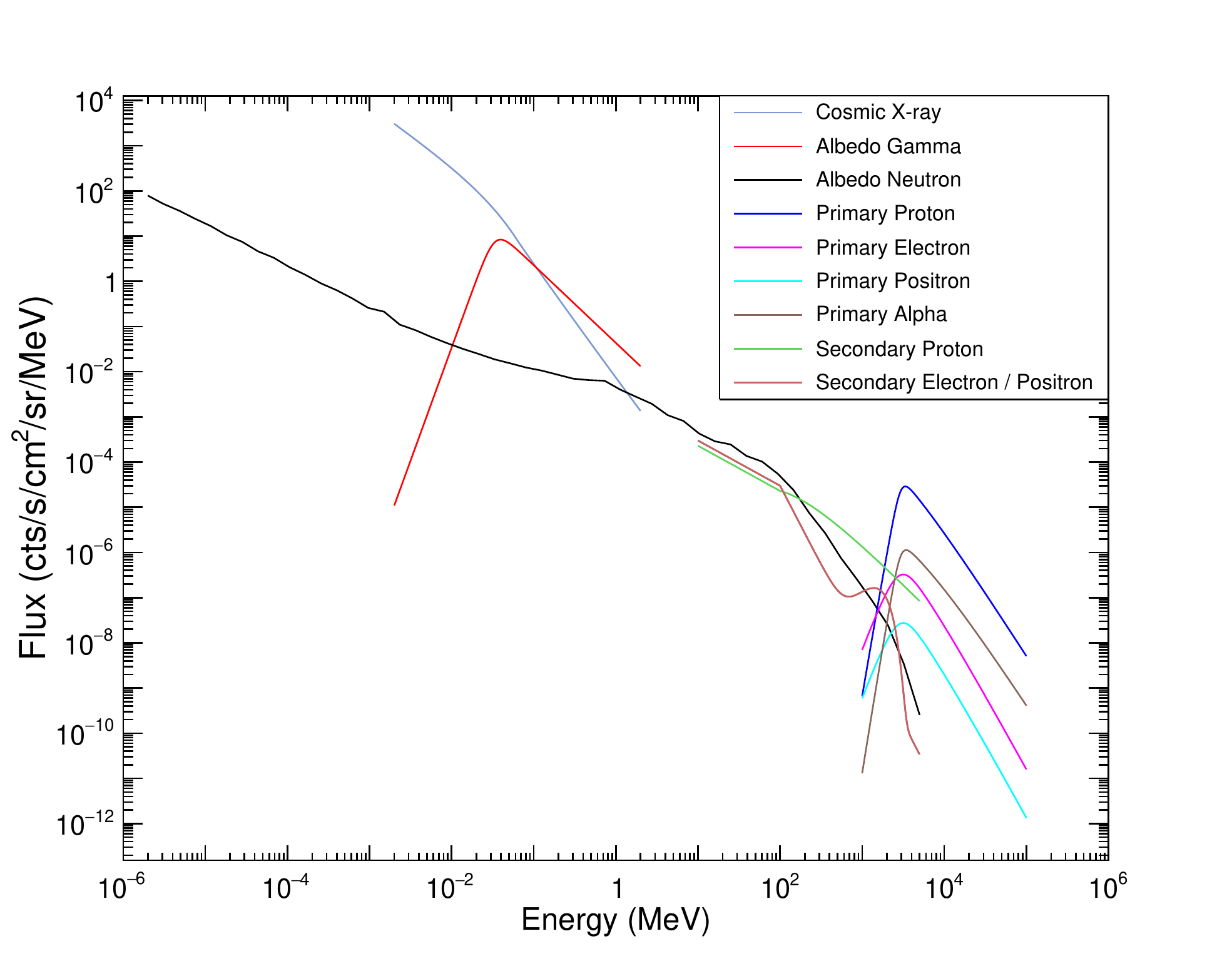}
\caption{The spectra of the ten background components simulated outside the SAA.} 
\label{fig:inputspec}
\end{figure}

The space radiation environment strongly depends on the orbit of the satellite. SPHiNX is foreseen to be launched into a low Earth orbit (LEO) with an altitude of 550~km and an inclination \mbox{of 53$^{\circ}$}. This~is the lowest available inclination due to platform constrains. Outside the South Atlantic Anomaly (SAA), prompt background contributions are mainly the cosmic X-ray background (CXB); primary and secondary cosmic rays (CRs), which are dominated by protons; albedo gamma rays; and neutrons, which are generated by the interaction of CRs with the upper atmosphere. 
\textcolor{black}{All background components are assumed to be isotropic. Earth-shielding effects are applied to CXB and cosmic-ray backgrounds. For albedo gamma rays and neutrons, the fluxes are normalised taking into account the average polarimeter pointing direction during a representative orbit.}
The energy spectra of these components are derived from the same methods used in the background simulation of HXMT (the~Hard X-ray Modulation Telescope)~\cite{ref:HXMT}, which operates in LEO at the same altitude of SPHiNX but a lower inclination of 43$^{\circ}$. \textcolor{black}{The spectra of cosmic rays exhibit a strong dependence on geomagnetic latitude. Compared to HXMT, SPHiNX operates at a lower geomagnetic cut-off which primarily affects low-energy particle fluxes.}
The input spectra are shown in Figure~\ref{fig:inputspec}, with the ten components considered for SPHiNX.  In the relatively high magnetic latitude, the secondary electrons and positrons have the same energy spectrum~\cite{ref:Mizuno}.

\begin{figure}[H]
\centering
\subfloat[Proton]{
\includegraphics[width=0.49\textwidth]{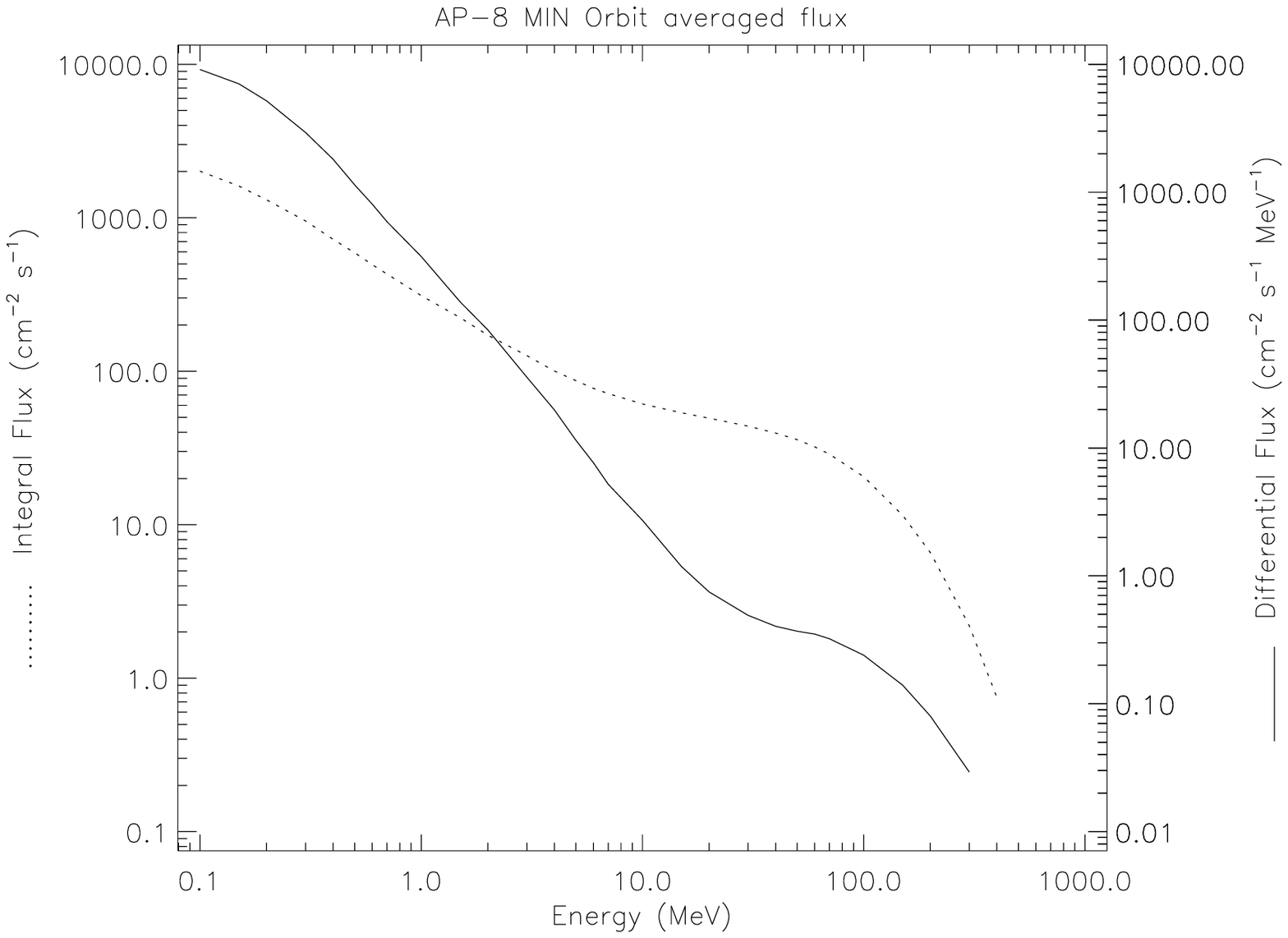}}
\hfill
\subfloat[Electron]{
\includegraphics[width=0.49\textwidth]{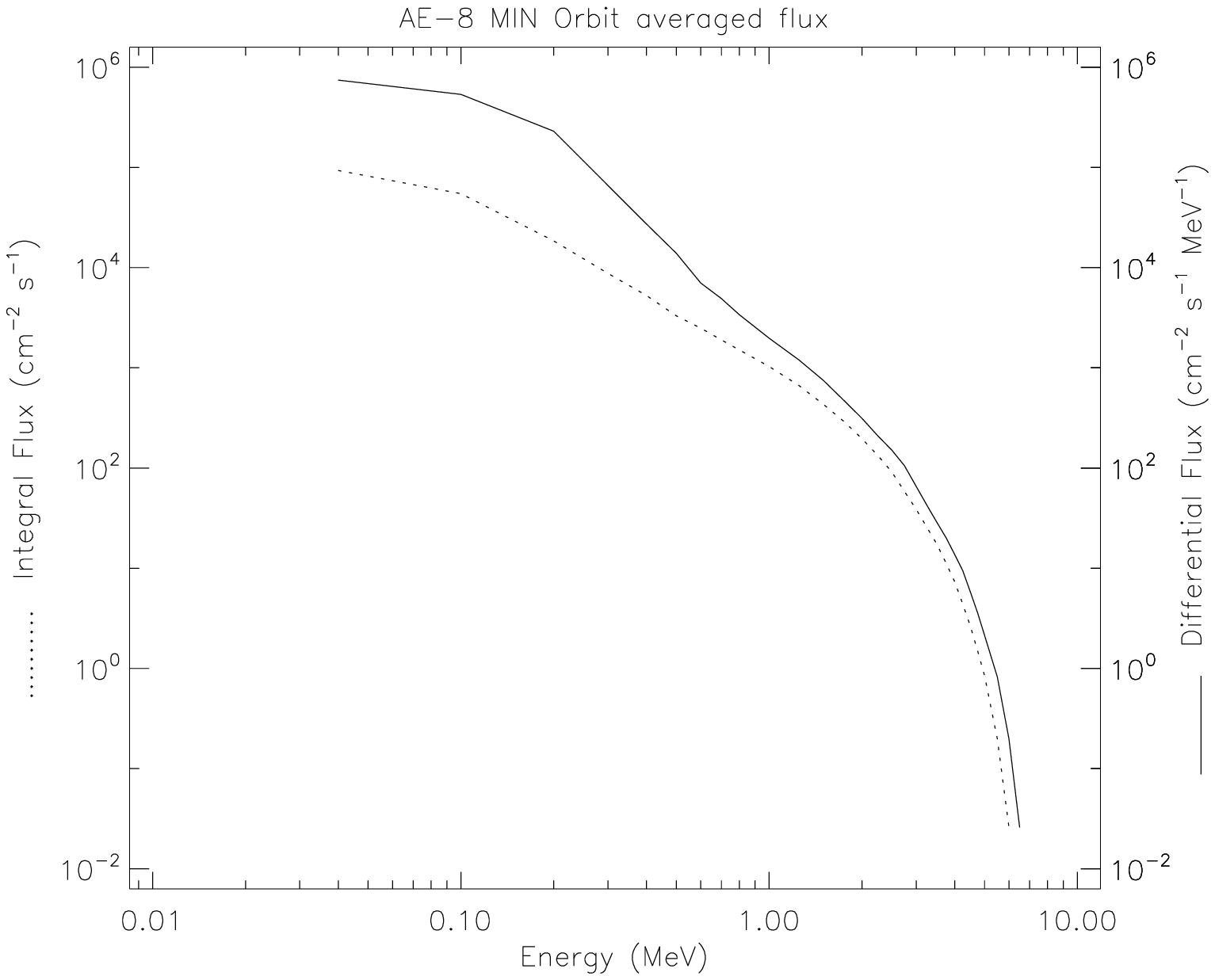}}
\captionsetup{width=1\linewidth}
\caption{Flux of SAA-trapped protons and electrons during solar minimum, as obtained from SPENVIS.} 
\label{fig:saaflux}
\end{figure}

\textcolor{black}{SPHiNX will pass through the SAA---a region with an extremely high flux of trapped particles (mainly protons and electrons)---several times per day.}
During a passage through the SAA, activation of structural components can cause a delayed background due to the decay of radioactive isotopes.
The spectra of trapped particles in the SAA are obtained from SPENVIS (ESA's SPace ENVironment Information System)\footnote{\url{https://www.spenvis.oma.be/}}. Prediction from models AP-8 and AE-8 for the orbital parameters of SPHiNX are shown in Figure~\ref{fig:saaflux}. The maximum galactic cosmic ray flux will arrive at the Earth during the solar minimum. SPHiNX is expected to be launched in 2021, which is approaching the solar maximum. Here, the conservative solar minimum case has been considered, corresponding to the extreme case that SPHiNX may encounter.
While the flux of electrons is much higher than the protons, the energy range of the electrons, which varies from 40 keV to 7 MeV as shown in Figure~\ref{fig:saaflux}b, is too small to activate materials. Protons, occupying a higher energy range up to 400 MeV, are the main source of the delayed background, as shown in Figure~\ref{fig:saaflux}a.

\unskip
\vspace{-12pt}

\subsection{Physics Model for Particle Interactions}
A reference physics list offered by the Geant4 collaboration named the Shielding Physics List has been used in the simulation. This package includes all the physics processes needed for simulations in the space environment, which include electromagnetic physics, hadronic physics and radioactive decay physics.
In order to correctly model the relatively low-energy interactions in the scintillators and treat polarisation properly, ``G4EmLivermorePolarizedPhysics'' has been used instead of the default ``G4EmStandardPhysics'' implementation.

\subsection{Data Analysis}

When tracking particles interacting with the sensitive detectors, all steps with energy deposit have been recorded together with the corresponding detector ID by Geant4 software. After the simulation of all the background components, the generated data are analysed.

For SPHiNX, an event will be considered as a signal only if certain photosensor pulse height conditions are fulfilled. Applying the same conditions to the background simulated data yields the background event rate. These conditions are related to three parameters, the hit threshold (HT), the~trigger threshold (TT) and the upper discriminator level (UD):
\begin{enumerate}[leftmargin=*, labelsep=4.9mm]
\item	All hits must have an energy deposit exceeding the HT.
\item	At least one hit has an energy deposit above the TT.
\item	No hit has an energy deposit exceeding the UD.
\end{enumerate}

Here, a hit means an interaction in a scintillator, which comprises the sum of all internal energy deposits. A hit is seen only when it has an energy deposit above the HT, otherwise it would be indistinguishable from the electronic noise level and would thus not issue a trigger in the instrument. TT is needed to flag a valid event and UD is applied for suppressing background, mainly from cosmic ray minimum-ionising particles (MIPs).

Based on the difference in light-yield between the plastic scintillator and GAGG scintillator, separate parameters have been applied for them, as shown in Table~\ref{table:cuts}. All these parameters are related to the detector and electronic readout system, and are chosen to optimise the polarimetric sensitivity in the SPHiNX energy range.

\begin{table}[H]
\caption{Parameters applied for plastic scintillator and GAGG scintillator.} 
\centering
\begin{tabular}{ccc}
\toprule
\textbf{Parameters}	& \textbf{Plastic (keV)}& \textbf{GAGG (keV)}\\
\midrule
HT	& 	5	&	30	\\
TT	& 	25	&	30	\\
UD	& 	600	&	600	\\
\bottomrule
\end{tabular}
\label{table:cuts}
\end{table}

\section{Results}
\label{sec:results}
\vspace{-6pt}

\subsection{Prompt Background}
The prompt background levels for all simulated background components outside the SAA after applying selections are shown in Table~\ref{table:bkgrate}. Here, the background count rate from each component has been separated into three categories: one-hit events (energy deposit in only one scintillator), two-hit events (energy deposit in two separate scintillators) and higher-multiplicity events (with interactions in three scintillators or more). Two-hit events are used to determine the polarisation, and one-hit events are foreseen to be used for spectroscopy and localisation of GRBs. Higher-multiplicity events may not be recorded due to onboard storage constraints.

From Table~\ref{table:bkgrate}, one-hit events are seen to dominate the total background at more than five times the two-hit events. The situation for GRBs is similar, with one-hit events dominating by percentages depending on the GRB energy spectrum. Only a fraction of one-hit events will be stored,  because of limitations on the onboard storage, dead time and downlink. Higher-multiplicity events contribute little to the total rate and are ignored in what follows.

The total prompt background of two-hit events is 323 counts/s, as shown in Table~\ref{table:bkgrate}. The~dominating component is CXB, with a contribution of 195 counts/s, which is $\sim$60\% of the total background. This is common for large-FoV satellite-borne detectors. The second prominent component is albedo gamma rays, coming from cosmic rays interacting with the upper atmosphere. The flux of albedo gamma rays depends on the amount of atmosphere that the detector views~\cite{ref:Dean}. Here, the average background from albedo gamma rays has been presented, which is about 35\% of the total background. The remaining 5\% comes from albedo neutrons and primary and secondary cosmic rays, which is negligible due to the shielding and thresholds applied, as discussed in Section~\ref{sec:ud}.

\begin{table}[H]
\caption{Background rates for all components.} 
\centering
\begin{tabular}{cccc}
\toprule
\textbf{Component}	& \textbf{One-Hit Rate (Hz)}& \textbf{Two-Hit Rate (Hz)}& \textbf{Higher-Multiplicity Rate (Hz)}\\
\midrule
CXB				& 	1270.1	&	195.3	&	37.5	\\
Albedo Gamma		& 	397.5	&	112.9	&	30.8	\\
Albedo Neutron		& 	14.3		&	5.1		&	2.7	\\
Primary CRs		& 	15.5		&	5.3		&	2.9	\\
Secondary CRs		& 	9.2		&	4.5		&	3.3	\\
\midrule
Total				&	1706.6	&	323.1	&	77.2	\\
\bottomrule
\end{tabular}
\label{table:bkgrate}
\end{table}

\subsection{Delayed Background}
The delayed background is generated by the decay of radioactive isotopes, produced by trapped protons in the energy range from 100~MeV to 400~MeV~\cite{ref:HXMT}. Such decays can be recognised from the energy deposit time recorded by Geant4. While the timescale for prompt background is $<$1~$\upmu$s, the~delayed background has a much wider time distribution, extending to hundreds of days depending on the half-life of the radioactive isotope.

\begin{figure}[H]
\centering
\includegraphics[width=0.8\textwidth]{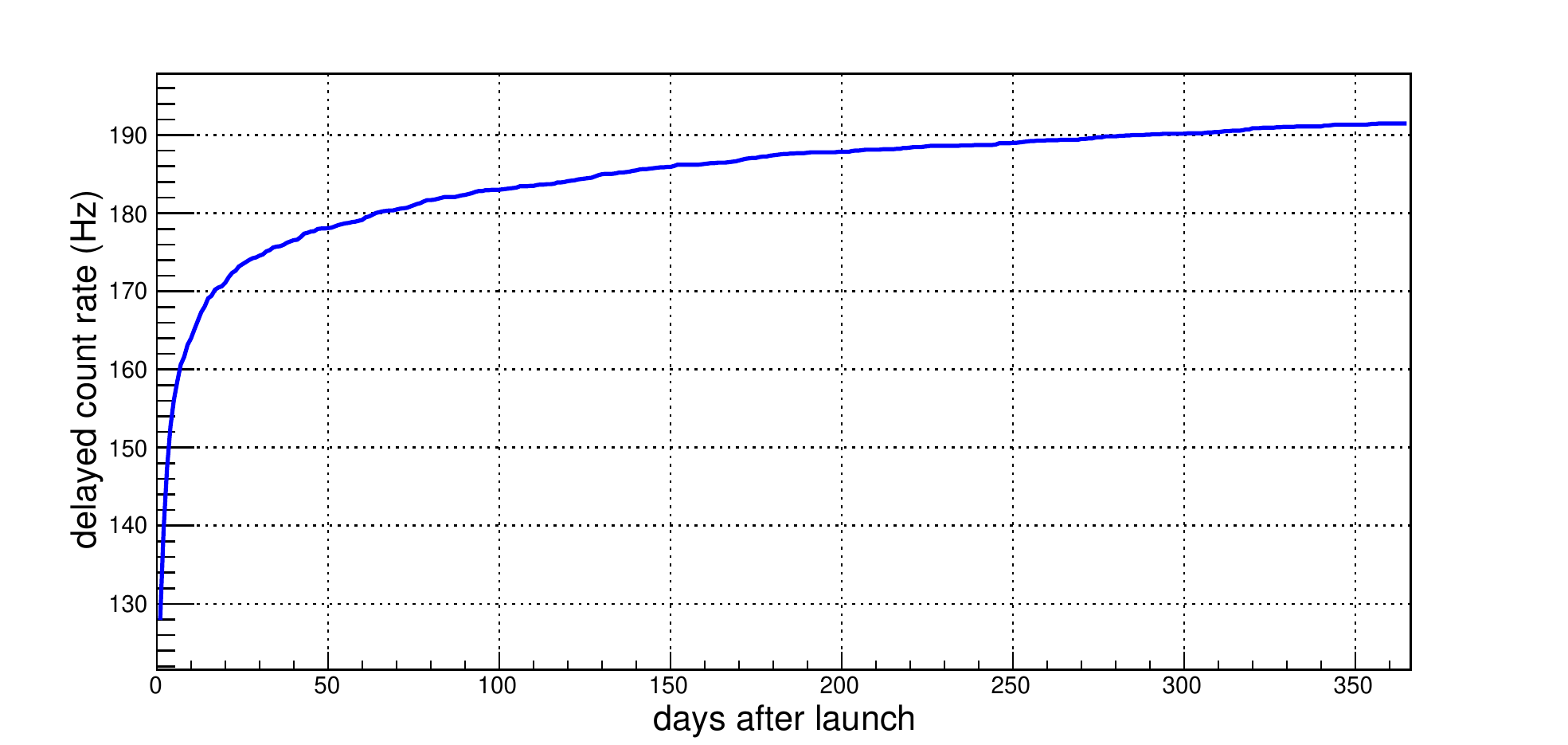}
\caption{The accumulated distribution of the delayed background of two-hit events induced by SAA-trapped protons.} 
\label{fig:saa}
\end{figure}

The simulation shows that the delayed background increases rapidly during the first month, and slowly saturates after one-year operation in orbit, to $\sim$190 counts/s for two-hit events, as  shown in Figure~\ref{fig:saa}.
The majority of the delayed background originates from the aluminium materials in the platform structure. The effect of this background may be reduced during final optimisation study of payload integration.
\vspace{-12pt}

\subsection{Upper Discriminator Selection}
\label{sec:ud}
The prompt and delayed background levels presented here are derived using a UD of 600~keV, which is chosen as a trade-off between source and background rates. The UD setting affects particle backgrounds (like protons, electrons and neutrons) more strongly than the photon background. For~example, the two-hit event rate from CXB increases by 0.7\% if no UD is applied at all, while~the primary proton rate increases by a factor of 22.5 in the absence of a UD selection, resulting in the two-hit event rate of $\sim$106 counts/s, which is comparable to the contribution from the second dominating background source, albedo gamma rays. As the UD selection only has limited influence on photons, \textcolor{black}{for a GRB with Band
spectral parameters~\cite{ref:Band}, $E_{peak} =$ 200~keV, $\alpha = -1.0, \beta = -2.5$}, UD~selection only affects 0.2\% of the two-hit event rate in maximum, which is negligible compared to the particle background. Thus, the UD selection is very important for the data quality in terms of the signal-to-background ratio. This is of high importance for SPHiNX, where the downlink capacity from the payload will be limited.

\subsection{Hit Threshold}
\begin{figure}[H]
\centering
\includegraphics[width=0.6\textwidth]{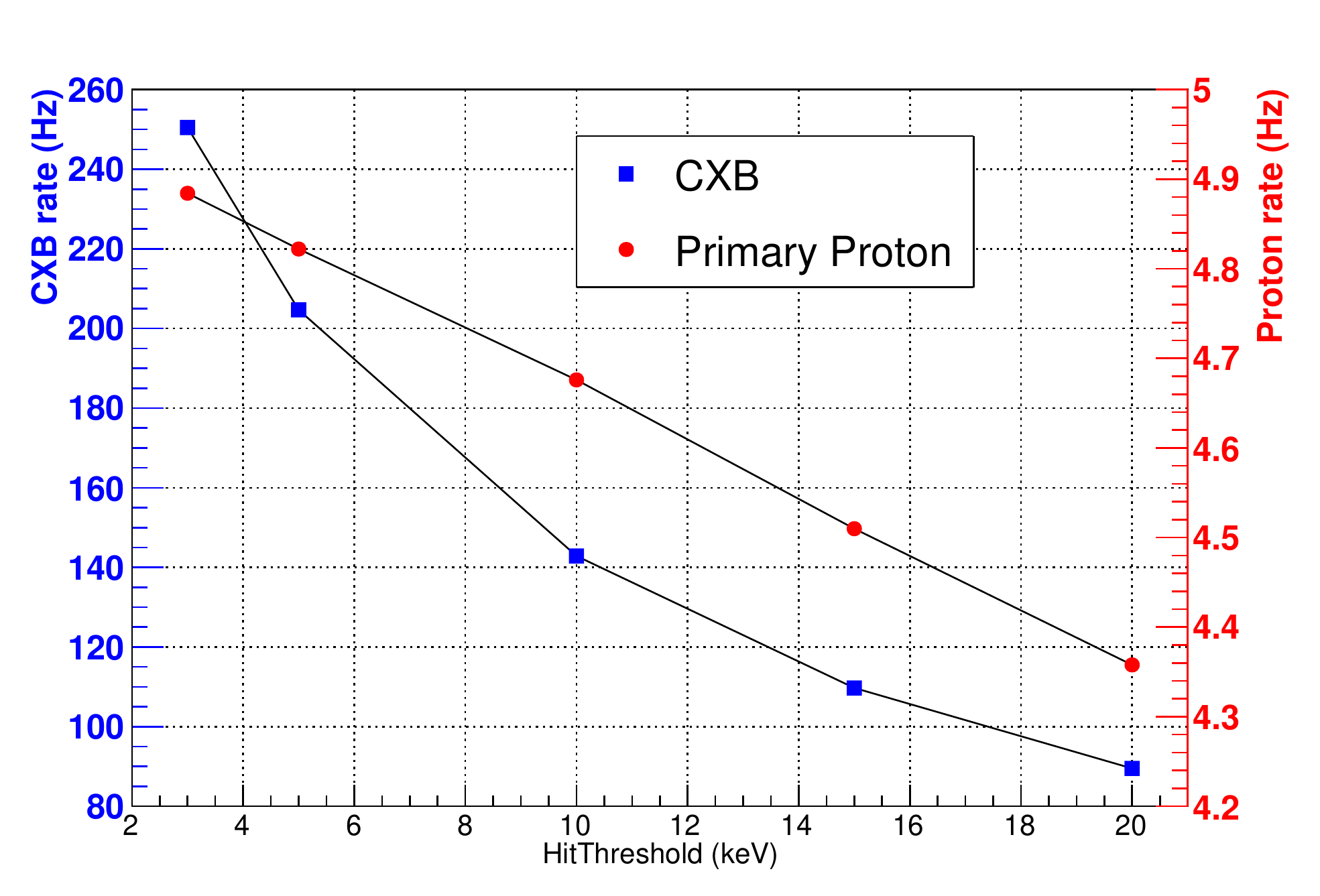}
\caption{Two-hit event rates of CXB and primary protons, as a function of hit threshold.}
\label{fig:ht}
\end{figure}

The behaviour of the HT is opposite to that of the UD. When the hit threshold is increased, the~CXB rate decreases dramatically, while the primary proton rate remains essentially unaffected, as shown in Figure~\ref{fig:ht}, where the same hit threshold is applied for both plastic and GAGG. This is under the assumption that the dynamic range of the GAGG detector and readout can be extended down to 5~keV. The total background is dominated by the rate of the CXB, that is, a high hit threshold is favoured. However, a large fraction of the signal events from the GRBs generate hits in this low-energy region, meaning that a low hit threshold is required for the scintillators. The optimisation of these event-selection parameters is done in terms of the minimum detectable polarisation (MDP)~\cite{ref:mdp} based on the Fermi/GBM catalogue of GRBs\textcolor{black}{~\cite{ref:sphinx}}.
\vspace{-12pt}

\section{Summary}
\label{sec:summary}
The Geant4 toolkit has been used in the background study of SPHiNX, by constructing an instrument model, a space radiation model and applying particle interaction physics models. Data~selections are based on the parameters listed in Table~\ref{table:cuts}, which are optimised in terms of the MDP based on an independent simulation of GRBs from the Fermi/GBM catalogue.

The simulation shows that the total background of two-hit events, including prompt and delayed background, is 513 counts/s when SHPiNX operates in orbit for one year. This is expected to be the most extreme case that SPHiNX will encounter.

The prompt two-hit event rate from all considered background components outside the SAA amounts to $\sim$323 counts/s, as shown in Table~\ref{table:bkgrate}. The dominant component is the CXB, contributing 60\% of the total background. The second most prominent component, albedo gamma rays, whose flux varies with the pointing of the instrument, has an average contribution of 35\%. Background from albedo neutrons and primary and secondary cosmic rays has a total contribution of 5\%, which is negligible, due to the shielding and thresholds applied.

A delayed background originating from radioactive isotope decay induced by SAA-trapped protons significantly increases the two-hit background. It increases rapidly during the first month, and~slowly saturates at $\sim$190 counts/s after SPHiNX is in orbit for one year, as shown in Figure~\ref{fig:saa}.
Since the delayed background mainly originates from the aluminium platform structure, it is expected that a reduction will be possible once the polarimeter shielding is optimised.

An independent simulation of GRBs from the Fermi/GBM catalogue, using the same thresholds as applied for the background study, shows that the polarisation of $\sim$50 GRBs with minimum detectable polarisation less than 30\% will be measured during the two-year mission lifetime.
The performance is sufficient to allow discrimination between GRB prompt emission arising from synchrotron processes in ordered and random magnetic fields, and inverse Compton-dominated outflows~\cite{ref:sphinx}. 

\vspace{6pt}

\authorcontributions{F.X. presented the paper at the Workshop on behalf of the SPHiNX Collaboration. \mbox{The manuscript} was prepared by F.X. and M.P.}

\acknowledgments{SPHiNX activities at KTH were funded by The Swedish National Space Board (SNSB). \mbox{OHB Sweden} are thanked for providing details of the InnoSat platform.}

\conflictsofinterest{The authors declare no conflict of interest.}


\reftitle{References}


\begin{thebibliography}{999}

\bibitem[Meszaros(2006)]{ref:grb}
Meszaros, P. Gamma-ray bursts. {\em Rep. Prog. Phys.} {\bf 2006}, {\em 69}, 2259. [\href{http://dx.doi.org/10.1088/0034-4885/69/8/R01}{CrossRef}]

\bibitem[Woosley(1993)]{ref:collapse}
Woosley, S.E. Gamma-ray bursts from stellar mass accretion disks around black holes. {\em Astrophys. J.} {\bf 1993}, {\em 405},~273--277. [\href{http://dx.doi.org/10.1086/172359}{CrossRef}]

\bibitem[Eichler(1989)]{ref:merge}
Eichler, D.; Livio, M.; Piran, T.;   Schramm, D.N. Nucleosynthesis, neutrino bursts and $\gamma$-rays from coalescing neutron stars. {\em Nature} {\bf 1989}, {\em 340}, 126--128. [\href{http://dx.doi.org/10.1038/340126a0}{CrossRef}]

\bibitem[Abbott(2017)]{ref:gw}
Abbott, B.P.; Abbott, R.; Abbott, T.D.; Acernese, F.; Ackley, K.; Adams, C.;  Affeldt, C. Gravitational waves and gamma-rays from a binary neutron star merger: GW170817 and GRB 170817A. {\em Astrophys. J. Lett.} {\bf 2017}, {\em 848}, L13. [\href{http://dx.doi.org/10.3847/2041-8213/aa920c}{CrossRef}]

\bibitem[Waxman(2003)]{ref:grbmodel}
Waxman, E. Astronomy: New direction for $\gamma$-rays. {\em Nature} {\bf 2003}, {\em 423}, 388--389. [\href{http://dx.doi.org/10.1038/423388a}{CrossRef}] [\href{http://www.ncbi.nlm.nih.gov/pubmed/12761528}{PubMed}]

\bibitem[Dean(2003)]{ref:Dean}
Dean, A.J.; Bird, A.J.; Diallo, N.; Ferguson, C.; Lockley, J.J.; Shaw, S.E.;  Willis, D.R. The modelling of background noise in astronomical gamma ray telescopes. {\em Space Sci. Rev.} {\bf 2003}, {\em 105}, 285--376. [\href{http://dx.doi.org/10.1023/A:1023995803108}{CrossRef}]

\bibitem[Agostinelli(2003)]{ref:geant4}
Agostinelli, S.; Allison, J.; Amako, K.A.; Apostolakis, J.; Araujo, H.; Arce, P.; Behner, F. GEANT4---A simulation toolkit. {\em Nucl. Instrum. Meth. A} {\bf 2003}, {\em 506}, 250--303. [\href{http://dx.doi.org/10.1016/S0168-9002(03)01368-8}{CrossRef}]

\bibitem[Produit(2018)]{ref:polar}
\textcolor{black}{
Produit, N.; Bao, T.W.; Batsch, T.; Bernasconi, T.; Britvich, I.; Cadoux, F.; Hajdas, W. Design and construction of the POLAR detector. {\em Nucl. Instrum. Meth. A} {\bf 2018}, {\em 877}, 259--268.
} [\href{http://dx.doi.org/10.1016/j.nima.2017.09.053}{CrossRef}]

\bibitem[Yoneyama(2018)]{ref:gagg}
\textcolor{black}{
Yoneyama, M.; Kataoka, J.; Arimoto, M.; Masuda, T.; Yoshino, M.; Kamada, K.; Usuki, Y. Evaluation of GAGG: Ce scintillators for future space applications. {\em J. Instrum.} {\bf 2018}, {\em 13}, P02023.
} [\href{http://dx.doi.org/10.1088/1748-0221/13/02/P02023}{CrossRef}]

\bibitem[Fei(2015)]{ref:HXMT}
Xie, F.; Zhang, J.; Song, L.M.; Xiong, S.L.; Guan, J. Simulation of the in-flight background for HXMT/HE. {\em Astrophys. Space. Sci.} {\bf 2015}, {\em 360}, 47. [\href{http://dx.doi.org/10.1007/s10509-015-2559-1}{CrossRef}]

\bibitem[Mizuno(2004)]{ref:Mizuno}
Mizuno, T.; Kamae, T.; Godfrey, G.; Handa, T.; Thompson, D.J.; Lauben, D.; Ozaki, M. Cosmic-ray background flux model based on a gamma-ray large area space telescope balloon flight engineering model. {\em Astrophys. J.} {\bf 2004}, {\em 614}, 1113--1123. [\href{http://dx.doi.org/10.1086/423801}{CrossRef}]

\bibitem[Band(1993)]{ref:Band}
Band, D.; Matteson, J.; Ford, L.; Schaefer, B.; Palmer, D.; Teegarden, B.; Fishman, G. BATSE observations of gamma-ray burst spectra. I-Spectral diversity. {\em Astrophys. J.} {\bf 1993}, {\em 413}, 281--292. [\href{http://dx.doi.org/10.1086/172995}{CrossRef}]

\bibitem[Weisskopf(2010)]{ref:mdp}
Weisskopf, M.C.; Elsner, R.F.; O'Dell, S.L. On understanding the figures of merit for detection and measurement of x-ray polarization. {\em Proc. SPIE} {\bf 2010}, {\em 7732}, 77320E. [\href{http://dx.doi.org/10.1117/12.857357}{CrossRef}]

\bibitem[Mark(2018)]{ref:sphinx}
Pearce, M.; Eliasson, L.; Iyer, N.K.; Kiss, M.; Kushwah, R.; Larsson, J.; Lundman, C.; Mikhalev, V.; Ryde, F.; Stana, T.A.; et al. Science prospects for SPHiNX---A small satellite GRB polarimetry mission. {\em Astropart. Phys.} {\bf 2019}, {\em 104}, 54--63. [\href{http://dx.doi.org/10.1016/j.astropartphys.2018.08.007}{CorssRef}]

\end{thebibliography}
\end{document}